\documentstyle[psfig]{mn}

\newif\ifAMStwofonts

\def\HII{H{\sevensize II}}

\ifoldfss
  \ifCUPmtlplainloaded \else
    \NewTextAlphabet{textbfit} {cmbxti10} {}
    \NewTextAlphabet{textbfss} {cmssbx10} {}
    \NewMathAlphabet{mathbfit} {cmbxti10} {} 
    \NewMathAlphabet{mathbfss} {cmssbx10} {} 
  \fi
  \ifAMStwofonts
    \ifCUPmtlplainloaded \else
      \NewSymbolFont{upmath} {eurm10}
      \NewSymbolFont{AMSa} {msam10}
      \NewMathSymbol{\upi}     {0}{upmath}{19}
      \NewMathSymbol{\umu}     {0}{upmath}{16}
      \NewMathSymbol{\upartial}{0}{upmath}{40}
      \NewMathSymbol{\leqslant}{3}{AMSa}{36}
      \NewMathSymbol{\geqslant}{3}{AMSa}{3E}

    \fi
  \fi
\fi 

\ifnfssone
  \newmathalphabet{\mathit}
  \addtoversion{normal}{\mathit}{cmr}{m}{it}
  \addtoversion{bold}{\mathit}{cmr}{bx}{it}
  \newmathalphabet{\mathbfit} 
  \addtoversion{normal}{\mathbfit}{cmr}{bx}{it}
  \addtoversion{bold}{\mathbfit}{cmr}{bx}{it}
  \newmathalphabet{\mathbfss} 
  \addtoversion{normal}{\mathbfss}{cmss}{bx}{n}
  \addtoversion{bold}{\mathbfss}{cmss}{bx}{n}
  \ifAMStwofonts
    \ifCUPmtlplainloaded \else
      \UseAMStwoboldmath
      \makeatletter
      \new@mathgroup\upmath@group
      \define@mathgroup\mv@normal\upmath@group{eur}{m}{n}
      \define@mathgroup\mv@bold\upmath@group{eur}{b}{n}
      \edef\UPM{\hexnumber\upmath@group}
      \new@mathgroup\amsa@group
      \define@mathgroup\mv@normal\amsa@group{msa}{m}{n}
      \define@mathgroup\mv@bold\amsa@group{msa}{m}{n}
      \edef\AMSa{\hexnumber\amsa@group}
      \makeatother
      \mathchardef\upi="0\UPM19
      \mathchardef\umu="0\UPM16
      \mathchardef\upartial="0\UPM40
      \mathchardef\leqslant="3\AMSa36
      \mathchardef\geqslant="3\AMSa3E
    \fi
  \fi
\fi 

\ifnfsstwo
  \DeclareMathAlphabet{\mathbfit}{OT1}{cmr}{bx}{it}
  \SetMathAlphabet\mathbfit{bold}{OT1}{cmr}{bx}{it}
  \DeclareMathAlphabet{\mathbfss}{OT1}{cmss}{bx}{n}
  \SetMathAlphabet\mathbfss{bold}{OT1}{cmss}{bx}{n}
  \ifAMStwofonts
    \ifCUPmtlplainloaded \else
      \DeclareSymbolFont{UPM}{U}{eur}{m}{n}
      \SetSymbolFont{UPM}{bold}{U}{eur}{b}{n}
      \DeclareSymbolFont{AMSa}{U}{msa}{m}{n}
      \DeclareMathSymbol{\upi}{0}{UPM}{"19}
      \DeclareMathSymbol{\umu}{0}{UPM}{"16}
      \DeclareMathSymbol{\upartial}{0}{UPM}{"40}
      \DeclareMathSymbol{\leqslant}{3}{AMSa}{"36}
     \DeclareMathSymbol{\geqslant}{3}{AMSa}{"3E}
    \fi
  \fi
\fi 

\ifCUPmtlplainloaded \else
  \ifAMStwofonts \else 
    \def\upi{\pi}
    \def\umu{\mu}
    \def\upartial{\partial}
  \fi
\fi

\title[Radiation hardening in galactic discs]{On the hardening of the ionizing radiation
in HII regions across galactic discs through softness parameters}
\author[E. P{\'e}rez-Montero \& J. M. V\'{\i}lchez]
       {Enrique P{\'e}rez-Montero$^{1}$ \& Jos\'e M. V\'\i lchez$^{1}$  \\
$^{1}$ Instituto de Astrof\'\i sica de Andaluc\'\i a. CSIC. Apartado de correos
3004. 18080, Granada, Spain. }
        
\date{Accepted 
      Received ;
      in original form November 2006}

\pagerange{\pageref{firstpage}--\pageref{lastpage}}
\pubyear{2006}

\begin{document}

\maketitle

\label{firstpage}

\begin{abstract}
We carry out a study of the hardness of the radiation of ionizing clusters in {\HII} regions.
Firstly, we explore the applicability of the softness parameter $\eta$, originally defined in the optical
for pairs of consecutive ionization stages of the same species. 
With the advent of the infrared space observatories, this definition has been extended to the 
mid-infrared.  We show that the softness parameters, as determined both in the optical and
the mid-infrared wavelengths,
are sensitive to the effective temperature using a sample of data in both spectral regimes. 
This is confirmed by comparing the data with a grid of photoionization models, although 
no complete agreement has been found even for different stellar model atmospheres.
Finally, we show that both softness parameters are consistent in the search for radial variations of 
the hardness of the ionizing radiation of {\HII} regions in
the discs of spiral galaxies. We find a range of trends, from galaxies showing pronounced gradients to
those showing very flat ones. Although the detectability and slope of these gradients can be altered by the 
size and luminosity of the studied {\HII} regions, it looks that their existence is related to the mass 
and type of the galaxies and, hence, to the properties of the entire disc.

\end{abstract}


\section{Introduction}

The radiation field emitted by massive stars ionizes the interstellar medium
producing the {\HII}  regions. These regions are found 
preferentially in gas-rich galaxy discs and star-forming dwarf galaxies and remain a
powerful tool to investigate many relevant
properties of the interstellar medium (ISM; metallicity, density, temperature) and
of the massive stellar populations of galaxies
(star formation rate, spectral energy distributions, IMF, among others). 

Since the pioneering works of Searle (1971)  the observation of
giant {\HII} regions in the discs of nearby spirals has 
revealed the existence of gradients of the excitation of the ionized gas, as
measured by the ratio of [O{\sc iii}]/H$\beta$ lines. These 
excitation gradients were originally associated to the existence of chemical
abundance gradients across the galaxies (Smith, 1975), nowadays a general property 
of spiral discs (e.g. Pilyugin et al., 2004).
In addition to the abundance gradient, it was originally found that a gradient in
the hardness of the ionizing spectral energy distribution (SED) 
was necessary in order to produce a detailed modeling of the spectra of (inner) HII
regions in some spiral galaxies (Shields \& Searle, 1978).
Indeed, the spectral sequence of (near zero-age/main sequence) giant {\HII} regions
has been shown to be driven by three main {\em functional} parameters: 
metallicity, {\em effective} ionizing  temperature  and ionization parameter
(e.g. McCall et al., 1985), allowing for a certain evolution of 
the ionizing stars and their nebulae (e.g.  Stasi{\'n}ska and Schaerer, 1997).


In order to put these earlier suggestions on a firmer observational basis, the
degeneracy between the spectral parameters excitation and metallicity must be broken, since 
the spectral observables of effective temperature, ionization parameter and 
metallicity are usually self-correlated. 
In this vein, V\'\i lchez \& Pagel (1988) defined a methodology to evaluate the
effective temperature of the ionizing stars in {\HII} regions independently (to first order) 
of the value of the ionization parameter; they used Osterbrock's equation for the abundance 
equilibrium of two consecutive ionization stages of two different species to produce a SED ``radiation 
softness'' parameter, thus removing the dependence on the ionization parameter:

\begin{equation}
\frac{n(X^{i+1})}{n(X^i)} = \frac{uc \langle\sigma^i\rangle}{\beta(X^i)} \cdot
\frac{Q(X^i)}{Q(H)}
\end{equation}

\noindent where $n(X^i)$ is the number density of the i-times ionized atoms of the
element $X$, $u$ is the ionization parameter, $c$ is the speed of the light, $\langle\sigma^i\rangle$ is the
mean value of the photoionization cross-section of $X^i$, $\beta(X^i)$ is the total
recombination coefficient, $Q(X^i)$ is the number of ionizing photons of the ion
$X^i$ and $Q(H)$ the number of ionizing photons of hydrogen. Since the quotient of ionizing photons is
sensitive to the shape of the ionizing SED and, hence, to the hardening of the radiation, this
quantity can be used to derive the effective temperature. The dependence on the ionization
parameter can be removed (to first order) by taking the ratio of two different quotients. This method can be used for
any pair of chosen consecutive ionization stages. Therefore, using the available
emission lines in the optical range of the {\HII} region spectrum, an optical ``radiation 
softness" parameter was defined as:

\begin{equation}
\eta = \frac{O^+/O^{2+}}{S^+/S^{2+}}
\end{equation}

\noindent which does not show a strong dependence on the electron temperature and it is proportional 
to the corresponding ratio based on the emission lines, 

\begin{equation}
\eta' = \frac{ I([O{\sc ii}] \lambda 3727)/I([O{\sc iii}] \lambda\lambda 4959,5007)
}{ I([S{\sc ii}] \lambda\lambda 6717,6731)/I([S{\sc iii}] \lambda\lambda 9069,9532)
}
\end{equation}

\noindent where all the wavelengths are in \AA . This
parameter results inversely related to the effective temperature, given the difference in 
the ionization potentials involved (O$^+$ IP 35.1 eV; S$^+$ IP 23.3 eV). Then, combining 
$\eta$ with one of the ionic quotients we can solve also for $u$.

More recently, Mart\'\i n-Hern\'andez et al. (2002) 
and  Morisset et al. (2004), among others,
have explored the use of similar softness parameters defined using the intensities of emission
lines available in the mid-infrared (mIR), in particular, based on 
Ar, Ne and S fine structure emission lines. Nevertheless, until recently it has not been possible
to carry out an empirical test of these mIR ``softness" parameters; presently this test is possible given 
the amount of data now available from space-based observations with {\em Spitzer Space Telescope}, 
in addition to previous data from the {\em Infrared Space Observatory}.

Spiral galaxies can offer the best laboratory to check, in a consistent way, the validity of these
mIR softness parameters, since the {\HII} regions in these discs have been observed under similar conditions 
and analyzed consistently. At present, there is a useful source of data available of
mIR emission lines of [Ne{\sc ii}], [Ne{\sc iii}], [S{\sc iii}] and [S{\sc iv}] in HII
regions of spiral discs (see Table 1). With this information, a ``softness" parameter in the mIR 
can be defined as follows:

\begin{equation}
\eta' (mIR) = \frac{ I([Ne{\sc ii}]\lambda 12.8\mu m)/I([Ne{\sc iii}]\lambda 15.6\mu m)
}{ I([S{\sc iii}]\lambda 18.7\mu m)/I([S{\sc iv}]\lambda 10.5\mu m) }
\end{equation}

To keep the negative correlation with the effective temperature --as in the definition of the optical parameter-- 
we used the line of Ne$^+$ (IP 41.0 eV) in the numerator over the line of S$^{2+}$ (IP 34.8 eV).
The study of the behaviour of this quotient using photoionization models (Morisset et al., 2004; 
Sim\'on-D\'\i az \& Stasi\'{n}ska, 2008) indicates a very slight dependence on the stellar atmosphere used in the 
models; and recent studies using a larger empirical database (cf. Groves et al., 2008) point to a clear offset of 
the AGNs when viewed within the same framework. 

Although recent evidences point towards possible variations in the ionizing SEDs of {\HII} regions 
across spiral discs, yet it remains unknown whether this behaviour appears an ``universal" feature of 
spiral galaxies. In particular, we do not know to which extent real gradients of the ionizing SEDs do exist in
spiral galaxies; whether these (possible) gradients of the ionizing SEDs could be parameterized in terms of the 
physical/chemical properties of spiral galaxies remains to be studied. 
With this aims, in this work we study empirically the applicability of both, optical and mIR softness 
parameters and make an exam of our results in the light of the predictions of a set of photoionization models. 
Then, using recent high quality data of {\HII} regions for a sample of spiral galaxies, we make a consistent study of 
the behaviour of the (gradients of) hardening of the ionizing radiation across their discs, as traced by the emission 
lines of their bright {\HII} regions. Finally, we discuss the results obtained and present our conclusions.

\section {Empirical "softness" parameters and photoionization models} 

We have made a selection from the recent literature of all high quality emission line fluxes 
necessary to calculate both, the optical and mIR softness parameters. In the case of the optical, 
we took the reddening-corrected spectroscopic data used in P\'erez-Montero et al. (2006),
with the addition of the {\HII} galaxies studied in H\"agele et al. (2008), a sample of 
metal-rich {\HII} regions in spiral discs (Bresolin et al., 2006) and the 
Circumnuclear Star Forming Regions (CNSFRs) studied in D\'\i az et al. (2007).

In the case of the mIR, we have compiled spectroscopic data from different sources
corresponding to {\em ISO} and {\em Spitzer} observations.  Table \ref{refs} lists the 
compiled objects along with their references and the source of observations. \begin{scriptsize}\begin{footnotesize}\begin{small}\begin{normalsize}\begin{large}\begin{Large}\begin{large}\begin{Large}\begin{LARGE}\begin{huge}\end{huge}\end{LARGE}\end{Large}\end{large}\end{Large}\end{large}\end{normalsize}\end{small}\end{footnotesize}\end{scriptsize} 


\begin{table}
\begin{minipage}{85mm}
\small
\vspace{-0.3cm}

\caption{Bibliographic references for the mid-IR emission line fluxes of the compiled sample
from Spitzer ($^a$) and ISO ($^b$).}
\begin{center}
\begin{tabular}{lcc}
\hline
\hline
Reference & Type or host & Pointings\\
\hline
Beir{\~a}o et al., 2008$^a$ & M82 & 6\\
Dale et al., 2009$^a$ & GEHRs & 149\\
Engelbracht et al., 2008$^a$ & BCDs & 41 \\
Giveon et al., 2002$^b$ & MW & 101 \\
Gordon et al., 2008$^a$ & M101 & 6 \\
Lebouteiller et al., 2008$^a$ & MW \& MCs & 31\\
Rubin et al., 2007$^a$ & M33 & 25\\
Rubin et al., 2008$^a$ & M83 & 23\\
Verma et al., 2003$^b$ & BCDs & 12\\
Vermeij et al., 2002$^b$ & MCs & 14\\
Wu et al., 2007$^a$ & BCDs & 11 \\
\hline
Total &  & 422 \\

\hline
\hline
\end{tabular}
\end{center}
\label{refs}
\end{minipage}
\end{table}



\begin{figure*}
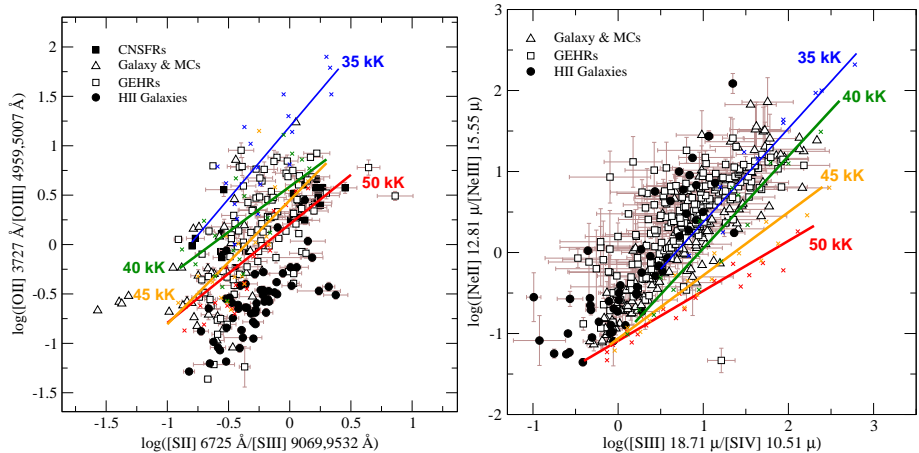

\begin{minipage}{170mm}
\centerline{
\psfig{figure=Letter_pmv09_fg01l.eps,width=6cm,clip=}
\psfig{figure=Letter_pmv09_fg01r.eps,width=6cm,clip=}}
\caption{Left panel: the ratio [S{\sc ii}]/[S{\sc iii}] versus [O{\sc ii}]/[O{\sc
iii}] in logarithm units for a sample of {\HII} regions,
HII galaxies and CNSFRs is compared with a grid of CLOUDY photoionization models.
The solid lines represent linear fits to the models with the same effective
temperature as indicated. Right panel: the mIR emission lines ratios 
[Ne{\sc ii}]/[Ne{\sc iii}] vs. [S{\sc iii}]/[S{\sc iv}] predicted by the same CLOUDY models
are compared with the corresponding flux measurements for the sample of objects with mid IR data.}
\label{eta}
\end{minipage}
\end{figure*} 

We make use of the photoionization code CLOUDY 08.00 (Ferland et al., 1998), taking as
ionizing source the SED of metal line-blanketed, NLTE, plane-parallel, hydrostatic model
atmospheres of O stars calculated with the code TLUSTY (Hubeny \& Lanz, 1995).
We assumed spherical geometry, a constant density of 100 particles per cm$^{3}$ and a
standard fraction of dust grains in the interstellar medium. We assumed as
well that the gas metallicity is the same for the ionizing source, covering the
values Z$_\odot$/30, Z$_\odot$/10, Z$_\odot$/5, Z$_\odot$/2 and Z$_\odot$,
taking as the solar metallicity the oxygen abundance measured by
Asplund et al. (2005; 12+log(O/H) = 8.66). The rest of ionic abundances have been set
to their solar proportions.
A certain amount of depletion was taken into account for the elements C, O, Mg, Si
and Fe, considered for the formation of dust grains in the code. 
Regarding other functional parameters we considered different values of the ionization
parameter (log U = -3.5, -3.0, -2.5 and -2.0) and of the stellar effective temperature
(T$_*$ = 35000 K, 40000 K, 45000 K and 50000 K). This gives a total of 80 photoionization 
models intended to cover the range of conditions of different ionized gas nebulae.


\section{Results and discussion}

In Figure 1 we show the relations between the ratios of emission lines used to
define both the optical and the mIR softness parameters.  
Both $\eta$'  are obtained directly by the lines of slope 1 with the same difference
between the {\em y} and the {\em x} axis. The models are represented by crosses and
the best linear fit to each set with the same  effective temperature are represented
as solid lines of different colours.

In the left panel, we show the plot of the optical softness parameter. The models fit
reasonably well the data, with the probable exception of {\HII} galaxies, which present abnormally high effective
temperatures as compared with the scale of the models; these values can be even higher 
than the stellar atmospheres with effective temperature of 50 kK. This feature has been
previously related to a (claimed) additional source of heating in these objects ({\em e.g}. Stasi\'{n}ska \&
Schaerer, 1999). 
At same time, we can observe
that higher effective temperatures correspond to lower values of the $\eta$'  parameter.
On the other hand, as was already found by Morisset (2004), lower metallicities lead to
lower values of $\eta'$, but this effect is sensibly lower in the considered range of metallicity than
those caused by differences in stellar effective temperature. Anyway it precludes to
provide a precise calibration of the effective temperature as a function of this parameter.

In the right panel, we show the equivalent plot for the emission line ratios of  the mid IR 
softness parameter. In this case, there is not a general agreement between the compiled
data and the predictions of the models for all the stellar temperatures; in such a way that a large fraction of the objects 
(mostly GEHRs) appears to be located in a region of the diagram corresponding to effective temperatures lower than 35 kK
This mismatch, already found by Morisset et al. (2004), does not look to be caused by the use of specific 
sets of stellar model atmospheres. For instance, the same grid of models but using the WMBasic stellar atmospheres 
(Pauldrach et al., 2002) produced offsets of 0.3 dex towards even lower values of $\eta$' .
Besides, models predict a similar dependence on metallicity as in the case of the optical parameter.

As in the case of the optical parameter, models also show a dependence of the mIR parameter on effective temperature,
in the sense that higher values of the mIR parameter are associated to lower model effective temperatures.
Nevertheless, the position of {\HII} galaxies in this diagram, overlapping a fraction of the data, and the confluence of 
the different slopes of the fits to the models towards the higher excitation corner of the plot, indicate that this parameter appears less efficient/reliable for those objects with higher ionization degree ({\em i.e.} lower [S{\sc iii}]/[S{\sc
iv}] ratios).


\begin{figure*}
\begin{minipage}{170mm}
\centerline{
\psfig{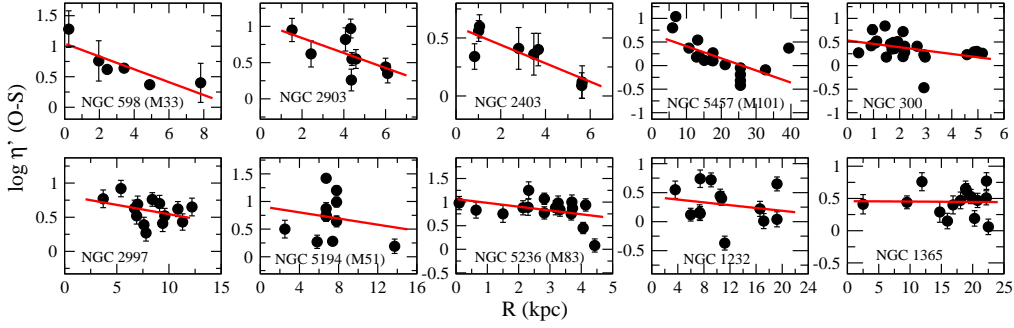}}
\caption{Deprojected radial variation of the optical softness parameter of {\HII} regions and corresponding best
linear fit for a sample of spiral discs. The slopes of the individual fits are listed in
Table \ref{slopes}. }
\label{grads_opt}
\end{minipage}
\end{figure*} 


\begin{figure*}
\begin{minipage}{170mm}
\centerline{
\psfig{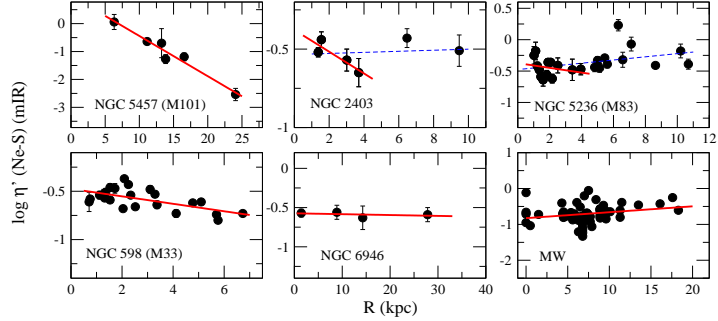}}
\caption{Deprojected radial variation of the mIR softness parameter of {\HII} regions and corresponding best linear
fit for a sample of spiral discs. For NGC2403 and NGC5236, solid lines show the fit to the common range with
the optical and dashed line the fit to all the data.}

\label{grads_mir}
\end{minipage}
\end{figure*} 


\begin{table}
\begin{minipage}{85mm}
\small
\vspace{-0.3cm}

\caption{Slopes of the deprojected gradients of both optical and mIR softness
parameters in a sample of spiral disc galaxies.}
\begin{center}
\begin{tabular}{lccl}
\hline
\hline
Galaxy name & log $\eta$' (opt) & log $\eta$' (mIR) & Ref.\footnote{B04: Bresolin et al., 2004; B05: Bresolin et al., 2005; B09: Bresolin et al., 2009; D08: Dale et al., 2009; G97, Garnett et al., 1997; G02, Giveon et al., 2002; G08: Gordon et al., 2008; K03: Kennicutt et al., 2003; R07: Rubin et al., 2007; R08: Rubin et al., 2008; V88: V\'\i lchez et al., 1988}
  \\
  & (dex/kpc) & (dex/kpc) \\
\hline
NGC 300   & -0.07$\pm$0.03 & -- & B09  \\
NGC 598   & -0.11$\pm$0.05 & -0.03$\pm$0.01 & V88, R07  \\
NGC 1232 & -0.01$\pm$0.02 & -- & B05 \\
NGC 1365 &  0.00$\pm$0.01 & -- & B05\\
NGC 2403 &  -0.08$\pm$0.02 & -0.07$\pm$0.02 & G97, D08 \\
NGC 2903 &  -0.10$\pm$0.05 & -- & B05 \\
NGC 2997 &  -0.03$\pm$0.02 & --  & B05 \\
NGC 5194 &  -0.03$\pm$0.05 & -- & B04 \\
NGC 5236 &  -0.03$\pm$0.02 & -0.04$\pm$0.05 & B05, R08  \\
NGC 5457 &  -0.08$\pm$0.05 & -0.14$\pm$0.02 & K03, G08 \\
NGC 6946  & -- & 0.00$\pm$0.01 & D08 \\
Milky Way & -- & 0.02$\pm$0.01 & G02 \\

\hline
\hline
\end{tabular}
\end{center}
\label{slopes}
\end{minipage}
\end{table}


The study of these softness parameters along the radial direction of the discs of
spiral galaxies provides a consistent scenario to study the variation of the hardening
of the ionizing radiation, mainly because these {\HII} regions have been observed consistently
across each galaxy.
In Figure 2, we show the deprojected gradient of the optical softness parameter for
sets of {\HII} regions in a sample of spiral discs for which the four required emission lines
have been measured.
In Table \ref{slopes}, we list the bibliographic sources from which we have extracted the
data for each galaxy as well as the slopes of the corresponding best linear fits, along with their errors. 
All of the galaxies show negative or flat radial gradients of $\eta$' (opt).  The values of the slopes 
ranging from 0.00$\pm$0.01 (NGC 1365) to -0.11$\pm$0.05 dex/kpc (NGC 598). This result 
indicates that, besides no radial positive gradient has been found, the existence of 
strong gradients of the hardening of the ionizing radiation in {\HII} regions of spirals {\em is not 
an universal property}. Rather, it seems that a range of slope values can be defined. 

In Figure 3, we show the gradients that were derived using the mIR softness parameter. Although 
a first impression gives us a large number of observations in this spectral regime, 
the poorer quality of a major part of the data --mainly due to low signal-to-noise
ratio of the [S{\sc iv}] emission line in {\HII} regions--, precludes this study 
in a high number of galaxies. The slopes of the linear fits and their errors are also listed
in Table \ref{slopes}.
As in the case of the optical parameter, we observe two main trends, with a pronounced gradient 
of the hardening of the radiation in some of the objects but a flat one in some others, as in the case 
of the Milky Way Galaxy. This behaviour we noted for our Galaxy was derived also by Morisset (2004)
using ISO data. 
A qualitative agreement is seen between both parameters
in those galaxies for which they can be measured simultaneously.
For instance, both parameters lead to pronounced gradients in M101 and within the common 
radii of NGC 2403 (the inner 5 kpc have both optical and mIR parameters). On the contrary, 
both parameters point to a somewhat flat gradients in M83. This agreement appears to reinforce 
the corresponding presence or absence of gradient in these galaxies. In the case of 
M33 the agreement between both slopes seems poorer, but still marginally consistent.
Unfortunately, given the limitations expressed above for the mIR and the lower number of galaxies analized, 
along this discussion we will use the slopes derived from this parameter only in a qualitative way.
 

\begin{figure}
\centerline{
\psfig{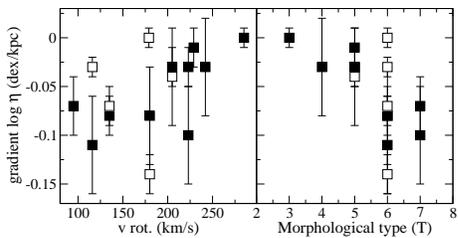}}

\caption{In left panel, relation between the rotation velocity of the spiral discs and the gradients of the
hardening of the ionizing radiation as obtained from the softness parameters. In right panel we show the relation between the gradient and the morphological type. Black squares represent the gradients obtained from the optical $\eta$' and white squares from
the mIR one.}
\label{corr}

\end{figure}


In order to try to understand the possible origin of the gradients derived for the ionizing SED of {\HII} regions 
in spiral galaxies, we have studied the correlation between the strength of these gradients and the properties 
of the disc {\HII} regions. In principle, one can expect that this relation could be driven by the existence of 
gradients of the metallicity across the discs. Though metallicity gradients have been derived for all these galaxies
giving a range of slopes, we failed to find a clear correlation between the slope of the 
metallicity gradient (taken from Pilyugin et al 2004) and the corresponding slopes of the gradients of the $\eta$' parameter 
This result does not exclude the possible correlation, within a given galaxy, between 
the softness parameter of {\HII} regions and their individual abundances (e.g. Vilchez and Pagel 1988; Bresolin et al 2009);
it is however clear that, at the galaxy scale, the strength of the $\eta$' gradient seems not to be
driven by the metallicity gradient.  
 
We have explored possible driving mechanisms of this behaviour at the galaxy scale. As we show in Figure \ref{corr} 
a significative trend appears to be present between the slopes of the gradients of the hardening the 
ionizing radiation --obtained with 
the optical softness parameter--, and the rotation velocity of the corresponding discs (taken from  
Pilyugin et al., 2004). This means that the gradient is more pronounced in the slow rotation, less massive 
galaxies. This fact may be pointing to a link between the evolutionary status of the galaxy discs and the 
properties across them of the ionizing radiation produced by the massive stellar clusters.
A hint in this direction can be seen in the right panel of Figure \ref{corr} where the slope of the 
$\eta$' gradient is shown versus the morphological type of galaxies. Again a trend is present in the figure, 
favouring stronger gradients for later type galaxies.
In this respect, it is necessary to bear in mind that the slopes of the gradients 
of the hardening of the ionizing radiation could be
driven by the star formation properties of the samples of {\HII} regions (e.g. efficiency of star formation) selected in the 
discs.
Overall, this fact would imply that faint/low surface brightness and bright {\HII} regions represent two families of objects 
that could reflect different stellar temperature behaviours across spiral discs. 
Thus, contrary to typical (bright) giant HII regions (more frequent in
later types), low surface brightness HII regions probably remain
less affected by the star formation conditions in the whole discs
(see Helmboldt et al.,  2009) .
In turns, this latter kind of HII regions seems more typical in earlier 
types.
This may be the paradigmatic case of the Milky Way, whose flat $\eta$'
gradient is traced mainly by small HII regions.

In summary, the existence of radial gradients of the hardening of the
radiation looks to be real and independently supported in
some galaxies by the results coming from the optical and the
mIR, but its relation with the mechanisms governing star
formation across the discs using HII regions of different
luminosities must be further investigated.

\section*{Acknowledgements}
This work has been supported by the project AYA-2007-08260-C03-02 of the Spanish
National Plan for Astronomy and Astrophysics. We thank P. Papaderos, G. Stasi\'{n}ska and C. Morisset for useful discussions.

\end{document}